\documentstyle[a4wide,epsf,11pt,titlepage]{article}
\newcounter{nref}
\newcommand{\bbib}{%
  \renewcommand{\refname}{\large\bf References}%
  \begin{thebibliography}{99}%
  \setcounter{enumiv}{\thenref}}
\newcommand{\ebib}{%
  \end{thebibliography}%
  \setcounter{nref}{\arabic{enumiv}}}
\newcommand{\head}[3]{%
  \setcounter{nref}{0}%
  \thispagestyle{empty}%
  \section*{\LARGE\bf #1}%
  \stepcounter{section}%
  \addcontentsline{toc}{section}{#1}%
  \large\itshape%
  #2\\\vspace{0.1pt}\\%
  #3%
  \normalsize\upshape%
  \bigskip}

\begin{document}


\head{Oscillations and convective motion in stars with \\ URCA shells}
     {G.S.\ Bisnovatyi-Kogan$^1$}
     {$^1$ Space Research Institute, Moscow\\}

\section*{Abstract}

It is shown that in presence of URCA shell pulsational energy
losses due to neutrino emission and
nonequilibrium beta heating are much less than
energy losses by excitation of short-wavelength acoustic waves.
Convective motion in presence of URCA shell is considered, and equations
generalizing the mean free path model of convection are derived.

\section{Introduction}

It was suggested in \cite{bkPaczynski1} that in the
convective region cooling of matter may be enhanced in presence
URCA shells
appearing when matter contains an isotope with a threshold Fermi energy for
an electron capture, coresponding to a density less then the central one.
Presence of such isotope leads to existence of a jump in the composition
at a density, corresponding to a threshold energy. During
convective motion the matter in eddies around this density crosses
periodically the boundary. That implies continious beta capture and beta
decay in the matter of these eddies.

Because of heating of a degenerate matter due to nonequilibrium
beta processes \cite{bkSeidov1}, with account of convective URCA
shell different conclusions had been done with respect to
stabilizing or destabilizing the carbon burning in the convective
degenerate core \cite{bkBruenn, bkErgma, bkPaczynski2, bkCouch,
bkLazareff, bkIben1, bkIben2, bkBarkat, bkMochkovitch, bkStein}.
Here we calculate damping of stellar oscillations in presence of
URCA shell (see \cite{bkIsern}), analyze physical processes in the
convective URCA shells and formulate approximate quantitative
approach to the solution of this problem.

\section{Linear oscillations of a slab with a phase transition}

Consider the plane-parallel layer in the constant gravitational
field with an acceleration $g$, with a phase transition at the pressure
$P_*$, and polytropic equation of state $P=K\rho^{\gamma}$,
with $K=K_1$ at $P<P_*$ and $K=K_2$ at $P>P_*$, $K_1>K_2$.
In static equilibrium
 $P=gM\left(1-\frac{m}{M}\right),\quad
 \rho=\left(\frac{gM}{K}\right)^{1/\gamma}
         \left(1-\frac{m}{M}\right)^{1/\gamma}.$
Here the pressure is continuous, but the density $\rho$ has a jump at $P_*$
due to the jump of the constant $K$, $M$ is the mass of one
cm$^2$ of the slab,
$m$ is the mass of one cm$^2$ of the slab under the layer with a Lagrangian
coordinate $x$, which is related to the density as
 $\rho=\left(\frac{\gamma-1}{\gamma K}\right)^{\frac{1}{\gamma-1}}
(C-gx)^{\frac{1}{\gamma-1}}$. In presence of a phase transition we
have
$ \rho_0=\left(\frac{gM}{K_2}\right)^{1/\gamma},\quad
 P_0=gM,\quad
 gx_*=C_2-{\frac{\gamma}{\gamma-1}}\frac{P_*}{\rho_{2*}}, \quad
 x_0=\frac{C_1}{g},\quad
 C_2=\frac{\gamma}{\gamma-1}(gM)^{\frac{\gamma-1}{\gamma}}K_2^{1/\gamma},\quad
 C_1=C_2+{\frac{\gamma}{\gamma-1}}\frac{P_*}{\rho_{2*}}(\lambda-1).$
Here $\rho_0$ and $P_0$ are the density and the pressure at the bottom
of the slab, $x_0$ and $x_*$ are total thickness and thickness of the
inner denser phase layer of the slab, $\lambda=\rho_{2*}/\rho_{1*}$.
The phase transition in the slab happens
only if its specific mass $M>P_*/g=M_*$.

Linear oscillations of the slab are reduced to Bessel equation for
perturbations ${\tilde P, \tilde v}
={\tilde P_a,\tilde v_a}\exp(-i\omega t)$ with solutions

\begin{equation}
\label{ref8}
 \tilde P_a=A{\sqrt z}J_{\frac{\gamma}{\gamma-1}}(\eta)+
          B{\sqrt z}Y_{\frac{\gamma}{\gamma-1}}(\eta),\quad
\end{equation}
\begin{equation}
\label{ref10}
 \tilde v_a= \frac{i}{\sqrt{\gamma P_0\rho_0}} z^{-\frac{1}{2\gamma}}
 \biggl[AJ_{\frac{1}{\gamma-1}}(\eta)+
          BY_{\frac{1}{\gamma-1}}(\eta)\biggr],
\end{equation}
where for two phases
$\eta_2=\frac{2\gamma}{\gamma-1}\frac{M\omega}{\sqrt{\gamma \rho_0
P_0}} z^{\frac{\gamma-1}{2\gamma}},\quad \eta_1=\eta_2
\sqrt{\lambda},\quad z_*=1-\frac{m}{M}$. The frequency of
oscillations $\omega$ and relations between constants
$A_1,\,\,B_1,\,\,A_2,\,\,B_2$ are obtained from boundary
conditions and from relations on the phase jump \cite{bkGrinfeld,
bkSeidov2, bkHaensel}. The dispersion equation are obtained
analytically for a frozen

\begin{equation}
\label{ref23}
J_{\frac{\gamma}{\gamma-1}}\left(\Omega\sqrt{\lambda}
  z_*^{\frac{\gamma-1}{2\gamma}}\right)\biggl[
  J_{\frac{1}{\gamma-1}}(\Omega)
  Y_{\frac{1}{\gamma-1}}\left(\Omega z_*^{\frac{\gamma-1}{2\gamma}}\right)
  -J_{\frac{1}{\gamma-1}}\left(\Omega z_*^{\frac{\gamma-1}{2\gamma}}\right)
  Y_{\frac{1}{\gamma-1}}(\Omega)\biggr]
\end{equation}
$$-\sqrt{\lambda}
J_{\frac{1}{\gamma-1}}\left(\Omega\sqrt{\lambda}
  z_*^{\frac{\gamma-1}{2\gamma}}\right)\biggl[
  J_{\frac{1}{\gamma-1}}(\Omega)
  Y_{\frac{\gamma}{\gamma-1}}
  \left(\Omega z_*^{\frac{\gamma-1}{2\gamma}}\right)
  -J_{\frac{\gamma}{\gamma-1}}
  \left(\Omega z_*^{\frac{\gamma-1}{2\gamma}}\right)
  Y_{\frac{1}{\gamma-1}}(\Omega)\biggr]=0$$
and equilibrium phase transition

$$J_{\frac{\gamma}{\gamma-1}}\left(\Omega \sqrt{\lambda}
  z_*^{\frac{\gamma-1}{2\gamma}}\right)\biggl[
  J_{\frac{1}{\gamma-1}}(\Omega)
  Y_{\frac{1}{\gamma-1}}\left(\Omega z_*^{\frac{\gamma-1}{2\gamma}}\right)
  -J_{\frac{1}{\gamma-1}}\left(\Omega z_*^{\frac{\gamma-1}{2\gamma}}\right)
  Y_{\frac{1}{\gamma-1}}(\Omega)\biggr]$$
\begin{equation}
\label{ref28}
-\biggl[\sqrt{\lambda}\,
J_{\frac{1}{\gamma-1}}\left(\Omega \sqrt{\lambda}
  z_*^{\frac{\gamma-1}{2\gamma}}\right)
   -(\lambda-1)\frac{\gamma-1}{2}\Omega
 z_*^{\frac{\gamma-1}{2\gamma}} J_{\frac{\gamma}{\gamma-1}}
 \left(\Omega \sqrt{\lambda}z_*^{\frac{\gamma-1}{2\gamma}}\right)
 \biggr]
\end{equation}
 $$\times \biggl[J_{\frac{1}{\gamma-1}}(\Omega)
  Y_{\frac{\gamma}{\gamma-1}}
  \left(\Omega z_*^{\frac{\gamma-1}{2\gamma}}\right)
  -J_{\frac{\gamma}{\gamma-1}}
  \left(\Omega z_*^{\frac{\gamma-1}{2\gamma}}\right)
  Y_{\frac{1}{\gamma-1}}(\Omega)\biggr]=0.$$
In the limiting case $m_*=0$, when the boundary between phases is on the
inner boundary of the layer, and $z_*=1$ the dispersion
equation is reduced to
$J_{\frac{1}{\gamma-1}}(\Omega \sqrt{\lambda})=0$
for a frozen, and

\begin{equation}
\label{ref29}
\sqrt{\lambda}\,
J_{\frac{1}{\gamma-1}}(\Omega \sqrt{\lambda})
   -(\lambda-1)\frac{\gamma-1}{2}\Omega
 J_{\frac{\gamma}{\gamma-1}}
 (\Omega \sqrt{\lambda})=0
\end{equation}
for an equilibrium phase transitions. At $m_* \rightarrow M, \,\,
z_* \rightarrow 0$ when the level between phases is moving to
the outer boundary, we have
the dispesion equation $J_{\frac{1}{\gamma-1}}(\Omega)=0$
in both cases. Here

\begin{equation}
\label{ref22}
\Omega
=\frac{2\gamma}{\gamma-1}\frac{M\omega}{\sqrt{\gamma \rho_0 P_0}},
\end{equation}
To investigate the dependence of oscillattion modes with an ideal
phase transition on $\gamma$ it is convenient to introduce
$\tilde\Omega=\frac{\gamma-1}{2\sqrt{\gamma}}\Omega$, with the
first root $\tilde\Omega^2 \rightarrow \frac{1}{\lambda-1}$ at
$\gamma \rightarrow \infty$. All other roots tend to infinity at
$\gamma \rightarrow \infty$. First three roots of equations
(\ref{ref23}), (\ref{ref28}) are presented in
Figs.1,2\footnote{Numerical solution of dispersion equations had
been done by O.V.Shorokhov.}.

\medskip

  {\bf Caption to Figures 1,2,3}

\noindent
Frequencies $\Omega$ of the slab oscillations as functions of $z_*$ of the
  basic mode (Fig.1), modes with one (Fig.2) and two (Fig.3) nodes. Upper and
lower curves correspond to frozen and equilibrium cases, relatively.
  Frequencies of these two cases coincide, when one of the node coincides with
  the phase transition.

\section{Damping of oscillations due to URCA
        shell in a highly degenerate matter}

Consider ultrarelativistic degenerate electron gas a good
approximation in most URCA shells.
It corresponds to the polytrope with $\gamma=4/3$, and constant
$K=\frac{c \hbar}{12 \pi^2} \left(\frac{3\pi^2}{\mu_Z m_p}\right)^{4/3}$,
where $\mu_Z=\left(x_Z\frac{Z}{A}+x_{Z-1}\frac{Z-1}{A}\right)^{-1}$ is the
average number of nucleons on one electron. Here a two component
mixture  is considered consisting of elements with an atomic weight $A$
and atomic numbers $Z$ and $Z-1$, with a beta transitions between them,
$x_Z$ and $x_{Z-1}$ are mass concentrations of these elements,
$x_Z+x_{Z-1}=1$.

Let $u_{fe}$ and $\delta$ be Fermi energy plus
rest mass energy of the electrons, and
threshold energy for a beta capture,
in units of $m_e c^2$; $g_z$ and $g_{Z-1}$ be
statistical weights of the elements $(A,Z)$ and $(A,Z-1)$; $Ft_{1/2}$
be a nondimensional value measured in the beta-decay experiments,
or estimated theoretically.
For small difference $\vert\delta-u_{fe} \vert \ll 1$ we have simple
expressions \cite{Bisnov} for
an entropy increase during
beta decay and capture in a fully degenerate matter

\begin{equation}
\label{ref34b}
\rho T \frac{\partial S}{\partial t}
=\Phi(\delta-u_{fe})^4 n_{Z-1},\quad
\rho T \frac{\partial S}{\partial t}=
=\Phi(u_{fe}-\delta)^4 \frac{g_{Z-1}}{g_Z} n_Z, \quad
\Phi=m_e c^2 \ln 2 \frac{(\delta^2-1)^{1/2} \delta}{12(Ft_{1/2})_{Z-1}}.
\end{equation}
Here the rate of the
entropy increase is equal to 1/3 of the energy loss rate by neutrino
emission.
During linear oscillations the beta reactions take place only in a thin
layer of matter, crossing in its motion the boundary
$x=x_*$, $m=m_*$, $P=P_*$.
In this layer the pressure may be represented by an expansion

\begin{equation}
\label{ref39}
P=P_*-g(m-m_*)+\tilde P, \qquad P^{1/4}= P_*^{1/4}+
\frac{1}{4} P_*^{-3/4}[\tilde P-g(m-m_*)],
\end{equation}
Equations describing change of concentrations during oscillations
averaged over the layer $\Delta m=\tilde P_a/g$, are written as

\begin{equation}
\label{ref43}
\frac{d \bar x_Z}{d t}=
-R\frac{P_*^{-9/4}}{256 }\tilde P_a^3 \cos^4{\omega t}
\frac{g_{Z-1}}{g_Z},
\qquad P_{eq}<P_*,\quad P>P_*;
\end{equation}
 $$
\frac{d\bar x_{Z-1}}{d t}=
-R\frac{P_*^{-9/4}}{256}
\tilde P_a^3 \cos^4{\omega t},\qquad P_{eq}>P_*,\quad P<P_*.$$
Here
 $R= \ln 2 \frac{(\delta^2-1)^{1/2} \delta}{3(Ft_{1/2})_{Z-1}}
   \left(\frac{ 12 \pi^2\hbar^3 }{m_e^4 c^5}\right)^{3/4}.$
To derive an equation describing decrease of the pulsational amplitude,
we should take into account the change of pressure due to change of
the electron concentration. Let $\tilde p$ be a perturbation of the
pressure determined by the beta reactions
$P=g(M-m)+\tilde P +\tilde p$, with
$\tilde p = \frac{4}{3} P \frac{\tilde x_Z}{Z},\quad P_{eq}<P_*,\,\,P>P_*;
\qquad \tilde p = -\frac{4}{3} P \frac{\tilde x_{Z-1}}{Z-1},
\qquad P_{eq}>P_*,\,\,P<P_*.$

In presence of damping we represent the velocity perturbation in a form
$\tilde v=\tilde v_a(t) \sin{\omega t}$,
and define $V(t)=\tilde v_{a,out}(m_*,t)$. so that
$\tilde v=V\sin{\omega t}$.
The amplitude of the perturbed outside pressure my be written as a
function of $V$ as
$ \tilde P_{a}=z_*^{\frac{7}{8}}
 \frac{\sqrt{\gamma P_0\rho_0}}{\sqrt{\lambda}}
\frac{J_4(\eta_{1*})}{J_3(\eta_{1*})} V,\quad \gamma=\frac{4}{3}$.
The equation of motion gives the relation describing damping of
oscillations in the layer $\Delta m(t) =\frac{\tilde P_a \cos
{\omega t}}{g}$, after subtraction of the proper oscillations of
the slab at the frozen composition and averaging over the motion
of the whole slab

\begin{equation}
\label{ref52}
\sin {\omega t}\frac{dV}{dt}=
-\frac{\Delta \tilde p}{\Delta m(t)} \frac{\Delta m}{M}
 \approx \frac{g\tilde p
(m_*)}{\tilde P_{a}\cos{\omega t}}\frac{\Delta m}{M},
\end{equation}
where $\Delta m=\frac{\tilde P_a}{g}$. Taking into account,
$M=gP_0$, we get

\begin{equation}
\label{ref53}
\sin {\omega t}\frac{dV}{dt}=\frac{g\tilde p(m_*)}{P_0\cos{\omega t}}.
\end{equation}
Averaging over the oscillation period we obtain equation for decreasing
of the amplitude of oscillations in the form

\begin{equation}
\label{ref57}
\frac{dV}{dt}=-D\,V^3,\qquad
D=\frac{R g P_*^{-5/4}}{768\omega P_0 Z }
\left(1+\frac{g_{Z-1}}{g_Z}\right)
(\gamma\lambda P_0\rho_0)^{3/2}
z_*^{\frac{21}{8}} \frac{J_4^3(\eta_{1*})}{J_3^3(\eta_{1*})}.
\end{equation}

\section{Energy balance and damping of oscillations}

Averaging equations (\ref{ref34b})
over the time and space we get an expression for heating rate of
the oscillating slab
due to nonequilibrium beta processes in the layer around the URCA shell

\begin{equation}
\label{ref45t}
\dot Q_{\nu}
=R_1\frac{P_*^{-3}}{g\, 32\times 75 \pi}\tilde P_a^5
\left(1+\frac{g_{Z-1}}{g_Z}\right), \quad
R_1=\ln 2 \frac{(\delta^2-1)^{1/2} \delta}{12(Ft_{1/2})_{Z-1}Am_p }
   \frac{ 12 \pi^2\hbar^3 }{m_e^3 c^3}.
\end{equation}
The rate of the neutrino energy losses
$L_{\nu}$ (ergs/s/cm$^2$) is obtained by
averaging over a time of losses in the whole oscillating layer.
We get similar to (\ref{ref45t})

\begin{equation}
\label{ref45c}
L_{\nu}=
3R_1\frac{P_*^{-3}}{g\,32\times 75 \pi}\tilde P_a^5
\left(1+\frac{g_{Z-1}}{g_Z}\right).
\end{equation}
Similar dependence of neutrino energy losses during oscillations because
of URCA shell had been obtained by Tsuruta and Cameron (1970), who also
took (\ref{ref45c}) for the rate of loss of kinetic energy of oscillations.
In the approximation of strong degeneracy the matter will be heated during
oscillations with the rate (\ref{ref45t}), and neutrino luminosity is
determined by (\ref{ref45c}). The source of both kind of energy fluxes is
the pulsation energy of the slab, giving the rate of pulsation energy
losses directly connected with beta reactions as

\begin{equation}
\label{ref62}
\dot E_{pul}^{(\beta)}=-(\dot Q_{\nu}+L_{\nu}) = -\frac{R_1 P_*^{-3}}
{600 \pi g} \tilde P_a^5\left(1+\frac{g_{Z-1}}{g_Z}\right).
\end{equation}
Defining $E_{\rm pul} \approx \frac{1}{2}MV^2$, and using approximately
$V=\frac{\tilde P_a}{\sqrt{\gamma P_0 \rho_0}}$,
for URCA shell in the middle of the slab, we get from
(\ref{ref57}) the
equation for decreasing of pulsational energy, connected with
hydrodynamic processes, in the form

\begin{equation}
\label{ref63}
\dot E_{\rm pul}^{(\rm dyn)} =-\frac{R P_*^{-5/4}}
{768\,Z \omega \sqrt{\gamma P_0 \rho_0}} \tilde P_a^4(1+\frac{g_{Z-1}}{g_Z}).
\end{equation}
A ratio of damping rates of oscillation on $i$-th mode is
$\frac{\dot E_{\rm pul}^{(\beta)}}{\dot E_{\rm pul}^{(\rm dyn)}}
\approx \frac {\tilde P_a}{P_0} \frac{\omega_i}{\omega_1}.
$
It follows that the main source of damping of
oscillations in presence of URCA shell is connected not with the neutrino
emission / nonequilibrium heating, but with a dynamical action of the
nonequilibrium layer of the slab, where beta reactions take place. This action
leads to an exitation of short-wavelength acoustic waves with length
$l \sim \frac{\Delta m}{\rho_0} \ll x_0$. When the
wavelength of the excited eigen-mode approaches the thickness of the
nonequilibrium layer $\Delta x$, formed by oscillations, both mechanisms
of damping become comparable. For $l_i \sim \Delta x
=\frac{\tilde P_a}{g \rho_*}$ with account of relations
$\omega_1 \approx \frac{1}{x_0} \sqrt{\frac{P_0}{\rho_0}}$,
$x_0 \sim \frac{P_0}{g \rho_0}$,
$\omega_i \approx \frac{1}{l_i} \sqrt{\frac{P_0}{\rho_0}}$, and $\rho_*
\sim \rho_0$, we get
$\dot E_{\rm pul}^{(\beta)}/\dot E_{\rm pul}^{(\rm dyn)} \sim 1$.
Importance of the
dynamical damping of stellar oscillations in presence of URCA shell is
connected with non-linearity of weak interaction rates, and deviations from
eigen-oscillations under an action of weak interactions in this layer,
leading to excitation of acoustic waves.
The dissipation of pressure oscillations, connected with
excitation of sound waves is inherent to any kind
of dissipation, when the main term is nonlinear.
The linear mechanism,
connected with a conventional bulk viscosity, does not change the form
of the
eigenfunction of oscillations, and so no additional waves are excited.

\section{URCA shell in a convective motion}

It was concluded in \cite{bkBarkat} that nonequilibrium heating is
balanced by the change in convective flow, leading to the net cooling
due to convective URCA shell. Nine years later same authors \cite{bkStein}
changed their mind, concluding that "convective URCA process
can reduce the rate of heating by nuclear reactions but cannot result
in a net decrease in entropy, and hence in temperature, for a constant or
increasing density." This conclusion, as well as opposite one, made by
using thermodynamic relations only, seems to be not convincing.
Following this line, let us present two plausible scenarios, leading to
two opposite conclusions.

{\bf A}. Due to action of a nonlinear bulk  viscosity, the convection is
damping in the vicinity of the URCA shell, decreasing the convective
heat flux from the central part of the star. In the general heat
balance of the star it means that cooling become less effective, and
nuclear reactions become thermally unstable and lead to a nuclear explosion
earlier, than without a presence of the URCA shell. Nonequibrium heating
give additional heating, supporting the earlier nuclear explosion.

{\bf B}. Due to action of a nonlinear bulk  viscosity, the convection is
damping in the vicinity of the URCA shell, decreasing the convective
heat flux from the central part of the star. Due to local decrease of the
heat flux from the core the average temperature gradient increases, leading
finally to increase of the convective flux soon after entering an URCA
shell into a convective zone. If the increase of the convective flux
prevails the nonequilibrium heating in the URCA shell, the general heat
balance would be shifted to a larger temperature with
more effective cooling, and the boundary
of the thermal explosion would be postponed in time, if not eliminated.

I cannot choose between these two scenario without
construction of the numerical model taking into account all processes
mentioned above. In such a highly nonlinear
system, as a star with nuclear
reactions, neutrino losses, degeneracy, convection, and many feedback
influences it seems to be impossible to make a conclusion about
the direction of process under the action of addtional URCA shell,
basing only on thermodynamical ground.

Convective modes belong to $g$-mode family, in which the local pressure
perturbations are small, and could be neglected when the
convective velocity is much less than the velocity of the sound. In
this situation the sound wave dissipation of the convective modes imposed by
URCA shell is negligible.
The equations of stellar evolution in presence of URCA shell should take
into account the following physical processes

1. loss of enegry due to neutrino emission in the URCA shell

2. heating of the matter in the convective region around
URCA shell due to nonequilibrium beta processes.

3. Decrease of the convective velocity in the layer around the URCA shell
due to energy dissipation connected with the nonequilibrium beta processes.
Kinetic energy of the convection is the source of energy for neutrino losses
and nonequilibrium heating of the matter.
In the condition of static equilibrium only energy and heat transfer
equations should be modified. In the energy equation

\begin{equation}
\label{ref67}
T\frac{dS}{dt}=\frac{dE}{dt}-\frac{P}{\rho^2}\frac{d\rho}{dt}=
\epsilon_n-\epsilon_{\nu}+\epsilon_{\nu}^{CU}-\frac{1}{4\pi\rho r^2}
\frac{dL_r}{dr},
\end{equation}
in addition to other neutrino cooling processes $\epsilon_{\nu}$,
the new term $\epsilon_{\nu}^{CU}$ is connected with heating due
to nonequilibrium beta processes around the URCA shell. having in
mind strong degeneracy of electrons in this region. Neutrino
emission in nonequilibrium URCA processes is accompanied by
heating at high degeneracy, because the positive term $\sum\mu_i
dn$ exceed the energy carried away by neutrino (Bisnovatyi-Kogan
and Seidov, 1970). Convective motion, consisting of convective
vortexes around an URCA shell is a source of additional neutrino
energy losses, and of heating of the matter. This dissipation of
convective energy may be described in the same way, as
corresponding dissipation and heating during stellar pulsations.
Therefore we use for description of these processes the formulae
from the previous sections. If we accept that the pressure
difference in the convective vortex is about one half of the local
pressure, we use for the amplitude of pressure pulsations in
(\ref{ref45t}) and (\ref{ref63}) $\tilde
P_a=\frac{\alpha_p}{4}P_*$. Taking also approximately
$P_0=P_*,\,\, \rho_0=\rho_*,\,\,$ and $u_{fe}=\delta$ we get

\begin{equation}
\label{ref70}
\dot Q_{\nu}
  = \tilde R\frac{m_e c^2}{Am_p}\frac{\delta^8 \alpha_p^5}
  {g\, 2^{10} \times 75 \pi},\quad
 \tilde R= \ln 2 \frac{(\delta^2-1)^{1/2} \delta}{12(Ft_{1/2})_{Z-1}}
   \frac{m_e^4 c^5}{ 12 \pi^2\hbar^3 }\left(1+\frac{g_{Z-1}}{g_Z}\right).
\end{equation}
Equation (\ref{ref70}) is related to energy losses averaged over the
whole slab. In the convective motion the losses are localized in the
layer around the URCA shell radius $r_*$

\begin{equation}
\label{ref72}
r_*+l_{\rm conv}<r<r_*-l_{\rm conv},\quad
l_{\rm conv}=\alpha_p \frac{P}{\nabla P},
\end{equation}
here $l_{\rm conv}$ is taken from the mean free path model.
The local rate is obtained from (\ref{ref70}), if we take
into account that the whole heating is concentrated inside the layer
(\ref{ref72}). We get than

\begin{equation}
\label{ref73}
\epsilon_{\nu}^{CU}=\frac{\dot Q_{\nu}}{2\rho l_{\rm conv}}.
\end{equation}
Convective velocity suffers from additional damping due to URCA shell,
because kinetic energy of the convection is a source of both nonequilibrium
heating and of additional neutrino losses. The only relation of the
convetional mixing length model of the convetion (see e.g. \cite{Bisnov})
should be modified, determining the convective velocity with an
additional damping

\begin{equation}
\label{ref79}
\frac{1}{2} \rho v_{\rm conv}^2=-\frac{1}{8}(\Delta\nabla T)l^2\,
\left.\left(\frac{\partial\rho}{\partial T}\right)\right|_P \,g
-\frac{\dot E_{\rm conv}^{(\beta)}}{v_{\rm conv}},
\end{equation}
where $\dot E_{\rm conv}^{(\beta)}=4\,\dot Q_{\nu}$  is found from
(\ref{ref70}).
These relations may be applied for description of URCA shell
convection only for sufficiently strong convective motion, when the
first term in (\ref{ref79}) exceeds considerably the second one.
The equation (\ref{ref79}) has roots only when

\begin{equation}
\label{ref81}
\dot E_{\rm conv}^{(\beta)}< \frac{1}{8\sqrt{\rho}}\left[-\frac{1}{3}
(\Delta\nabla T)l^2\,
\left.\left(\frac{\partial\rho}{\partial T}\right)\right|_P
\,g\right]^{3/2}.
\end{equation}
Violation of this inequality may result in an abrupt termination
of the convection in the layer (\ref{ref72}) around the URCA
shell. When analysing the URCA shell convection in star, it would
be premature to predict the results of evolutionary calculations
with account of convective URCA shell before such calculations are
done. Two possibilities may be expected . One is connected with
obtaining of a definite result which has a little sensitivity to
the input parameters of the problem, such as $\alpha_p$,
$Ft_{1/2}$, accepted rates of nuclear reactions, neutrino losses
etc. Another possibility could be a great sensitivity of the
result to the same input parameters. If the second possibility
would be realized we could still remain in situation of ambiguity,
because the set of the input parameters for presupernovae model
cannot be established with a sufficient precision.

\section*{Acknowledgements}

Author is grateful to R.Canal, S.I.Blinnikov, J.Isern, R.Mochkovich
for useful discussions; to W.Hillebrandt and E.M\"uller for
the invitations and posssibility to participate in
workshops ''Nuclear Astrophysics', to O.V.Shorokhov for help.
This work was partly supported by Russian
Basic Research Foundation
grant No. 99-02-18180 and grant of a Ministry of Science and Technology
1.2.6.5.


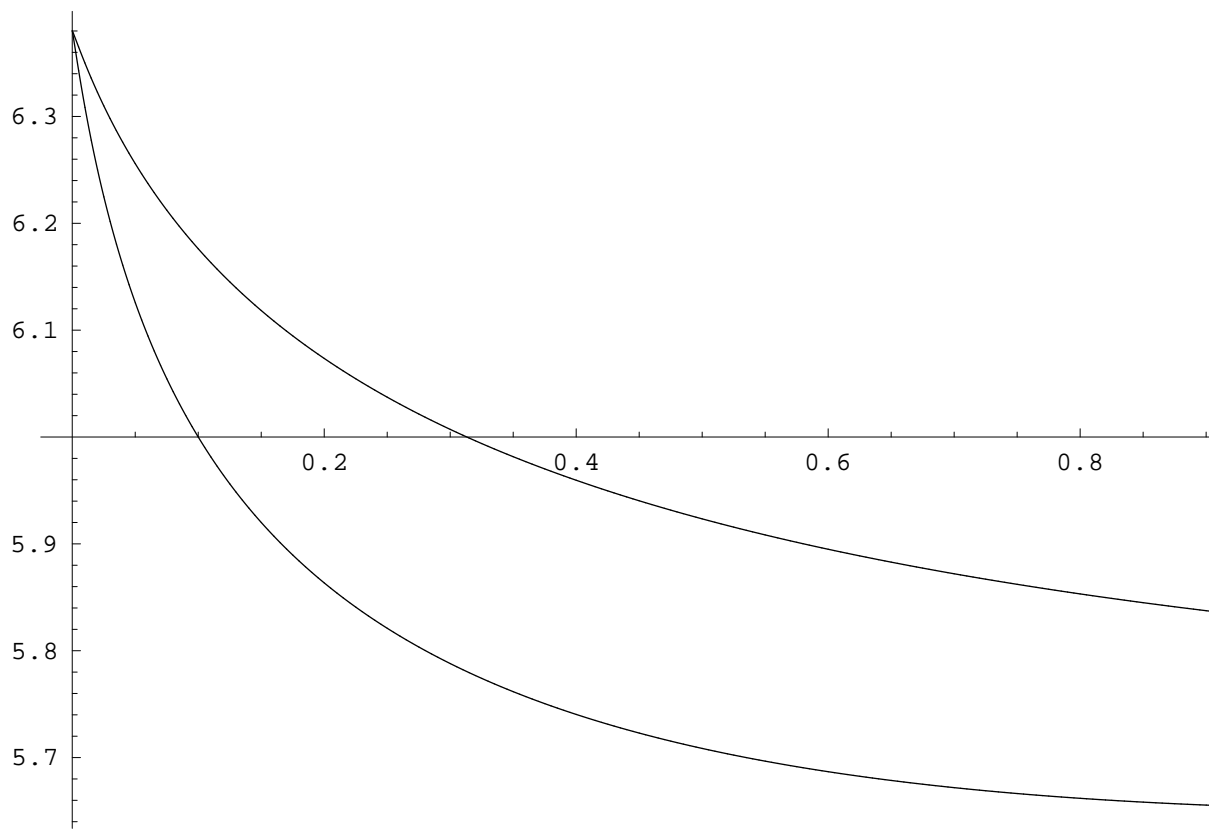
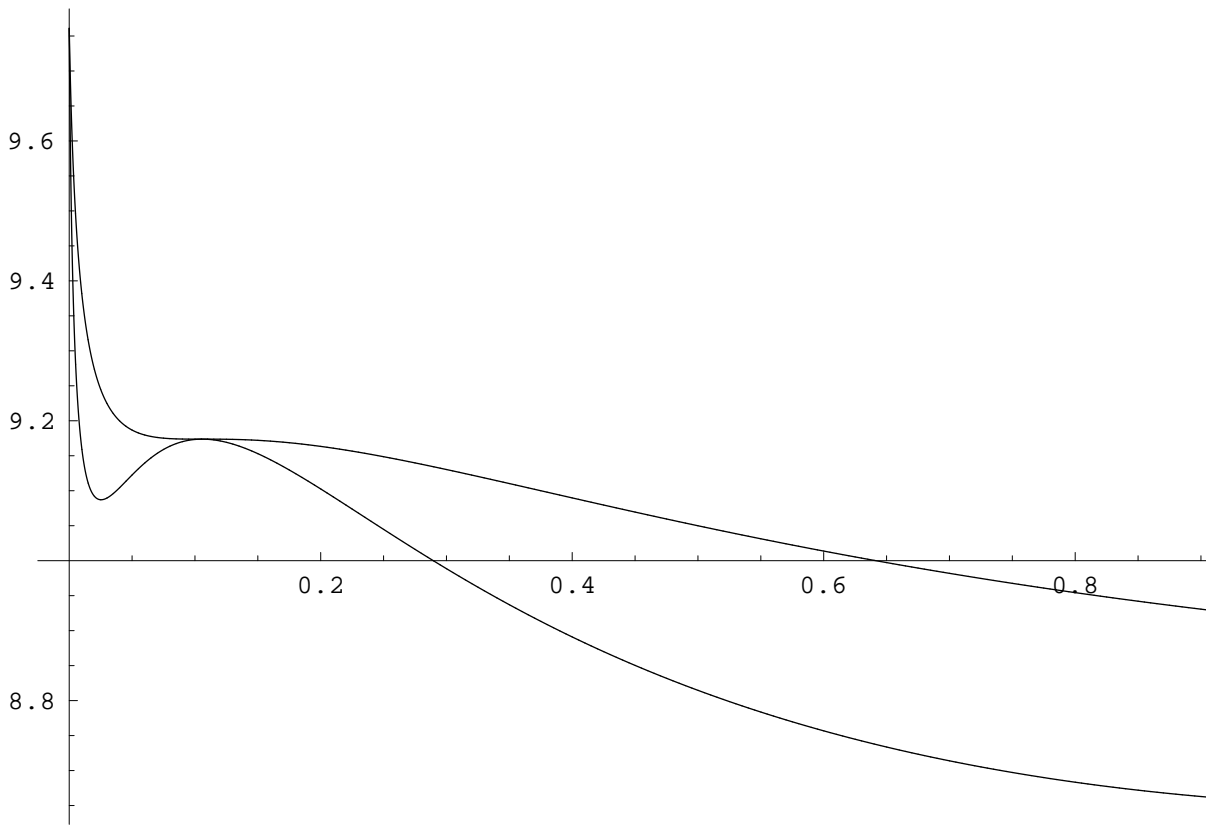
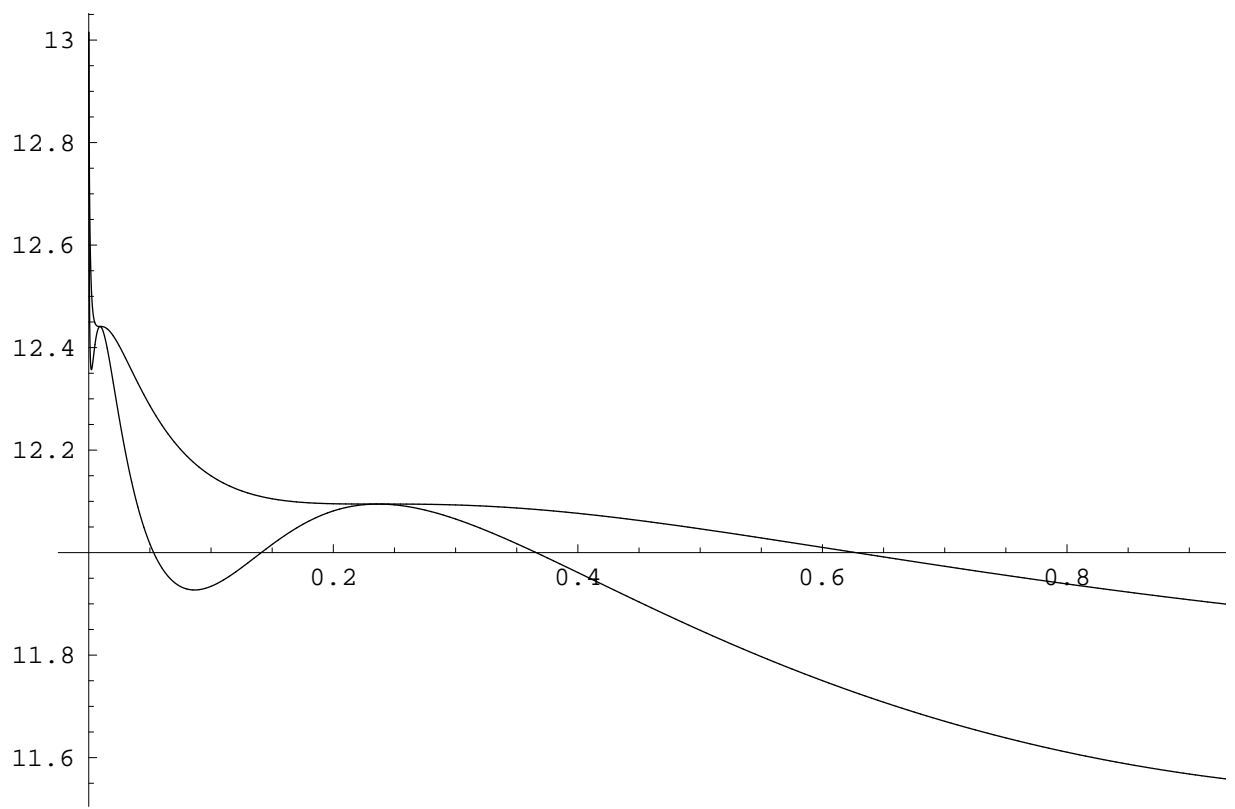
\end{document}